\documentclass[final,5p,twocolumn,times]{elsarticle}
\usepackage{url}

\usepackage{graphicx}

\usepackage{amssymb}

\newcommand{\GP}{\mbox{GALPROP}}
\newcommand{\WR}{\mbox{WebRun}}
\newcommand{\GPWR}{\GP~\WR}
\newcommand{\gray}{$\gamma$-ray}

\journal{Computer Physics Communication}

\begin{document}

\begin{frontmatter}



\title{\GP\ WebRun: an internet-based service for calculating \\ 
    galactic cosmic ray propagation and associated photon emissions}


\author{
A.~E.~Vladimirov,
S.~W.~Digel,
G.~J\'ohannesson,
P.~F.~Michelson, 
I.~V.~Moskalenko, 
P.~L.~Nolan,
E.~Orlando, 
T.~A.~Porter}
\address{W. W. Hansen Experimental Physics Laboratory, Kavli Institute for Particle Astrophysics and Cosmology, Department of Physics and SLAC National Accelerator Laboratory, Stanford University, Stanford, CA 94305, USA}


\author{A.~W.~Strong
}
\address{Max-Planck Institut f\"ur extraterrestrische Physik, 85748 Garching, Germany}

\begin{abstract}
\GP\ is a numerical code for calculating the galactic propagation 
of relativistic charged particles and the diffuse emissions
produced during their propagation. 
The code incorporates as much realistic astrophysical input as 
possible together with latest
theoretical developments and has become a {\it de facto} standard in 
astrophysics of cosmic rays. 
We present \GPWR, a service to the scientific community enabling easy 
use of the freely available \GP\ code via web browsers. 
In addition, 
we introduce the latest \GP\ version 54, available through this service.
\end{abstract}


%
%

\end{frontmatter}

\section{Introduction}

A large number of outstanding problems in physics and astrophysics
are connected with studies of cosmic rays (CRs) and the associated 
diffuse emissions 
(radio, microwave, X-rays, \gray{s}) produced during their propagation.
Among these problems are indirect searches for dark matter (DM), the origin
and propagation of CRs, particle acceleration in putative CR
sources -- such as supernova remnants -- and the interstellar 
medium (ISM),
Cosmic rays in other galaxies and the role they play in galactic evolution, 
studies of our local Galactic environment, CR propagation in the
heliosphere, the origin of the 511 keV line from the Galactic Center, the 
origin of the extragalactic diffuse emission, as well as many others.
New or improved instrumentation to explore these open issues are operating or 
under development.
Low-energy CR detectors on spacecraft and balloons, such as ACE, 
TIGER, the Voyagers 1 and 2, provide data on isotopic composition, 
and are complemented by the currently on-orbit PAMELA experiment, 
which is designed to measure antiprotons as well as light nuclei, electrons, 
and positrons above 100 GeV.
Elemental spectra are provided up to the TeV range by the 
experiments CREAM, ATIC, and TRACER, while PPB-BETS, CALET, and 
the {\it Fermi}--LAT provide measurements of electrons.
Instruments such as INTEGRAL, {\it Fermi}--LAT, HESS, MILAGRO, MAGIC, and 
VERITAS cover from hard X-rays to \gray{s} up to TeV
energies.
A number of other experiments are in the research phase, under construction, 
or about to be launched,
for example, AMS, OASIS, HAWK, and CREST. 
For a recent review see \cite{SMP07}.

The complex nature of these scientific goals, such as, e.g., the detection
of a weak DM signal on top of the intense diffuse \gray{} emission
produced by CRs interacting with the ISM, or the 
study of K-capture isotopes and electrons in CRs, 
requires \emph{reliable and detailed calculations} using a numerical model.
All of the latest developments of astrophysics, and 
particle and nuclear physics, play a role in addressing these questions: 
the latest developments in CR acceleration and transport mechanisms, 
detailed maps of the three-dimensional Galactic gas distribution, 
detailed studies of the interstellar dust, radiation field, 
and magnetic field, as well as the Local Bubble, and
new particle and nuclear cross section data and codes.
Achieving these scientific goals requires a realistic, yet simple to access
and use, model with most of the technical details left to experts.
Alternatively, sophisticated users must have access to the full source code and
documentation to easily allow them to modify the code as suits their purpose.

\begin{figure*}
   \centering
   \includegraphics[width=2\columnwidth]{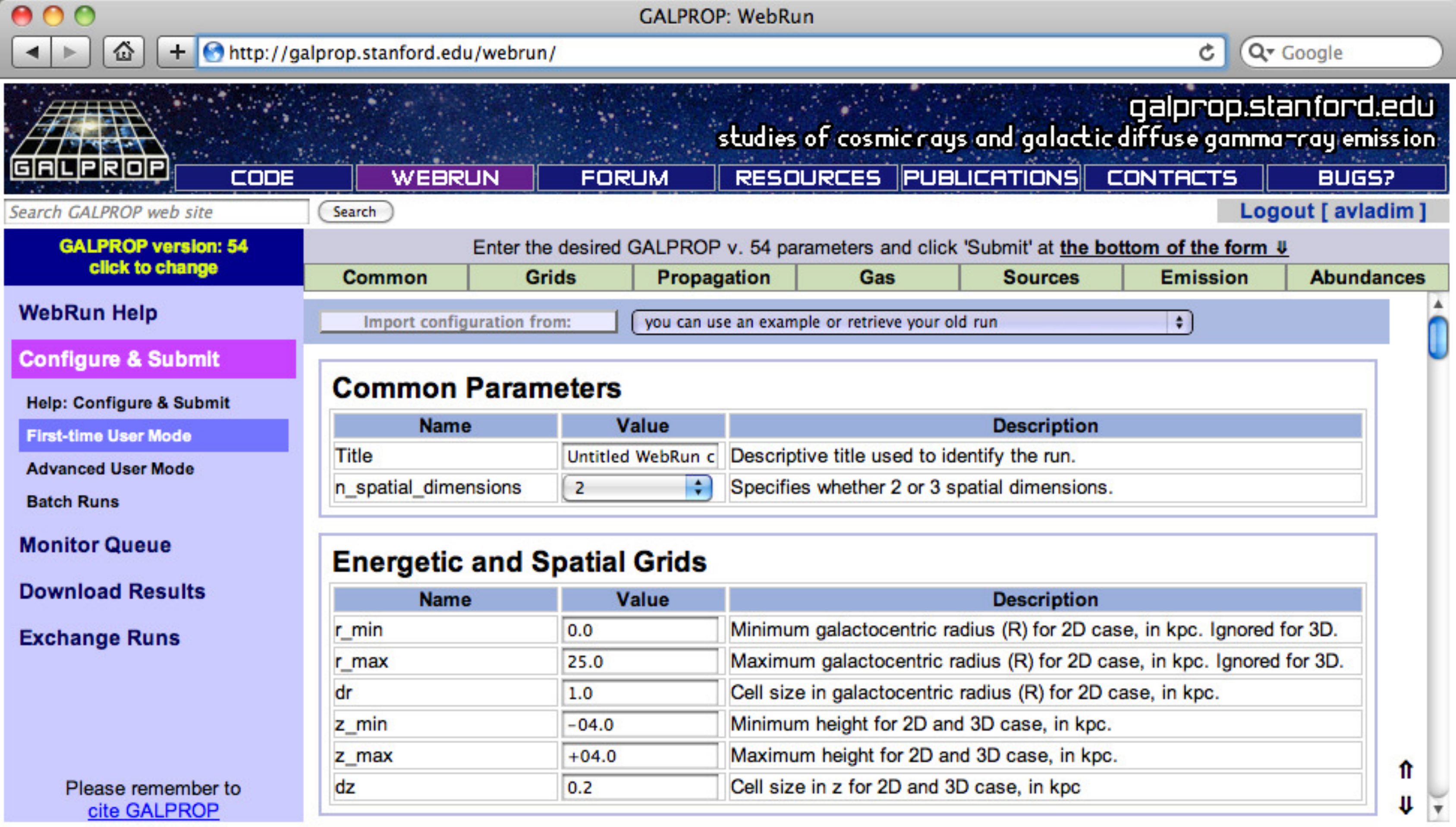}
   \caption{The \GPWR\ service allows users to create a \GP\ model 
     using a web browser, validate the model parameters to reduce the 
     likelihood of misconfigured runs, and execute the \GP\ code on a 
     dedicated computing cluster. 
   The service is accessible via http://galprop.stanford.edu/webrun/.}   
   \label{webrun}
\end{figure*}

\GP\ is a numerical code for calculating the propagation of relativistic 
charged particles and the diffuse emissions produced during their 
propagation. 
The \GP\ project has been running since the mid-1990s, and since then it has evolved into a sophisticated, efficient, configurable tool for high energy astrophysics \cite{wwwsciencewatch}.
The code incorporates as much realistic astrophysical input as possible 
together with latest theoretical developments. 
\GP\ calculates the propagation of CR nuclei, antiprotons, 
electrons and positrons, and computes diffuse \gray{s} and 
synchrotron emission in the same framework. 
Each run of the code is governed by a configuration file allowing 
the user to specify and control many details of the calculation. 
Thus, each run of the code corresponds to a potentially different ``model''. 
The code is written mainly in C++ along with some well-tested 
routines in Fortran 77.

Formalism and results of the \GP\ code can be found in 
papers 
\citep[][]{SM98, MS98, MS99, MS00, SMR00, MSOP02, MSMO03, MGMPS04, SMR04, SMR04b, SMRDD04, MSOM05, PMJSM05, MPS06, PMS06, PMJSZ06, PMSOB08, SPDJMMMO10, TJMPRS2010}, 
conference proceedings 
\citep[][]{2001ICRC....5.1836M, 2001ICRC....5.1868M, 2003ICRC....4.1917M, 2003ICRC....4.1921M, 2003ICRC....4.1969M, 2005AIPC..745..585S, 2005AIPC..769.1612M, 2005AIPC..801...57M, 2007AIPC..921..411P, 2008ICRC....2..129M, 2008ICRC....2..509H, 2008AIPC.1085..763A, 2009essu.confE..16B}, 
and reviews 
\citep[][]{SM01, MSR04, SMP07}.

Until now, the only way to use the \GP\ code has been downloading the 
source code from the project web site, installing and running it on a 
user's machine. 
Besides occasional problems with the installation process, users often 
experienced difficulties configuring the \GP\ runs.

\section{The GALPROP WebRun service}

The \GPWR\ service presented here enables any member of the 
scientific community to use the most recent version of \GP\ via a web browser. 
Installation of the code locally is not required with the calculations 
performed on a dedicated computing cluster employing AMD Opteron 6174 CPUs
managed by the \GP\ team 
(cluster specifications are given in Section~\ref{cluster}).
In addition, using the input web forms of \WR\ 
allows the input parameters to have sanity checks applied via rules 
written by the \GP\ team, thus minimizing the risk of misconfigured \GP\ 
runs (see Figure~\ref{webrun} for an example).
Each user has a queue that can be used for batch jobs of 
multiple \GP\ runs, where the user load is dynamically distributed 
across the available compute nodes.
We also provide the latest stable release of \GP\ along with earlier tagged 
releases to allow cross-checking of results across different versions.

The results of a \GP\ run (CR distributions and diffuse 
emission sky maps) are 
written out in FITS files readable by common astronomical software
\cite{wwwnasafits}. 
Some of this data (e.g., energy spectra of CR isotopes, abundances of CR 
species, and spectra of diffuse emissions) can be plotted using the \WR\ 
interface. 
The output of a \GP\ run generally contains much more usable information 
than given by the built-in plot templates in \GPWR, so users are 
encouraged to build their own plotting routines that use the data files 
produced by \GP.

\WR\ stores the results of each run along with the \GP\ configuration 
file on the web server from where the author of the results can download them 
as a compressed `tar' archive. 
Sharing the results is also made easy by \WR: the author of the run can 
publish the URL of the tar archive, which will make it possible for anyone to 
download and peruse the results.
The URLs of user-generated \WR\ archives are encoded with a string of 16 random characters. This string acts as a key to the run known only to the user that generated it, unless he or she decides to share or publish the URL. 

The \GPWR\ service can be accessed on the 
World Wide Web
\cite{wwwwebrun}.
Registration is required to use \WR, which is done via a form at the \WR\ URL.

\section{New features in the latest GALPROP~version~54}

Simultaneously with the release of the \WR\ service, we present an 
improved version the \GP\ code (v54). 
This version is available to the scientific community only 
via the \GPWR\ site. 
The new features of the code include:
\begin{itemize}

\item
Shared-memory parallel support with OpenMP to take advantage of 
multi-processor machines;

\item
Memory usage optimization;

\item
Implementation of the HEALPix output of gamma-ray and synchrotron 
skymaps. 
The HEALPix format \citep{G05} is a standard for radioastronomy 
applications, as well as for such instruments as WMAP, Planck etc.;

\item
Implementation of the MapCube output for compatibility 
with Fermi-LAT Science Tools software
\cite{wwwfermilat};

\item
Implementation of gamma-ray skymaps output in Galactocentric 
rings to facilitate spatial analysis of the Galactic diffuse gamma-ray 
emission;

\item
More accurate line-of-sight integration for computing 
diffuse emission skymaps;

\item
3D modeling of the Galactic magnetic field, both regular and 
random components, with a range of models from the literature, 
extensible to any new model as required;

\item
Calculations of synchrotron skymaps on a frequency grid, using 
both regular and random magnetic fields;

\item
Improved gas maps, which are computed using 
recent H~I and CO (H$_2$ tracer) surveys \citep[][]{Kalberla2005, Dame2001}, with more precise assignment 
to Galactocentric rings;

\item
A new calculation of the Galactic interstellar radiation field 
using the FRaNKIE code (Fast Radiation transport Numerical Kode for 
Interstellar Emission, as described in \citep{PMSOB08}) and implementation 
of the corresponding changes in GALPROP;

\item
Considerably increased efficiency of anisotropic inverse 
Compton scattering calculations;

\item
\GP\ code is compiled to a library for easy linking with other 
codes (e.g., DarkSUSY \cite{wwwdarksusy}, SuperBayeS \cite{wwwsuperbayes});

\item
Numerous bug fixes and code-style improvements;

\item
Improved configuration management via the GNU autotools. Multiple 
*NIX system and compiler targets (gcc, intel, llvm, open64) are supported;

\item
Bugzilla available at the \GPWR\ URL for user-submitted bug tracking.

\end{itemize}

\section{Computing cluster specifications}
\label{cluster}

\WR\ runs on a cluster comprised of a head node, which is used as a web and
data server, and four compute nodes that are connected with Gb ethernet and
Infiniband links.
The head node is 2-way machine using 12-core AMD Opteron 6174 processors 
with 64 GB of physical memory and 16 TB of redundant data storage.
Each compute node is a 4-way system using 12-core AMD Opteron 6174 processors
with 128 GB of physical memory and a modest amount of local storage via 
redundant disks ($2\times300$ GB mirrored drives) 
for the operating system and other critical files.
Data sets and other common files are shared via NFS from the head node, as is 
the common storage area that run results from the compute nodes are written to. 
Since the current GALPROP parallelization scheme uses a shared memory model 
(via OpenMP), these AMD-based systems have enabled us to consolidate 
computational and memory resources within a single unit to enable even 
high-resolution 3D calculations.
In addition, these systems allow us to easily extend the parallel
calculations to CPUs and add-in GPU cards using, e.g., 
OpenCL, or even a 
hybrid MPI/OpenMP/OpenCL scheme (with some rearchitecting) 
to use more than one compute node in the future.
Similar capabilities were not possible with, e.g., a blade-based system given 
our power and cost-per-unit requirements.
At any rate, 
this relatively modest system is easily extended dependent on demand for
the \WR\ service.

The \GPWR\ cluster uses the CentOS
\cite{wwwcentos}
Linux distribution along 
with the OSCAR
\cite{wwwoscar}
software suite that is used to
install, configure, and synchronize, the software across the cluster.
The run queue and resource allocation for user runs of \GP\ via \WR\ are
managed using TORQUE
\cite{wwwtorque}
in combination with
Maui
\cite{wwwmaui}.
User registration and authentication in \WR\ is linked
to a database on the \GP\ forum, which based on the
open source bulletin board software phpBB
\cite{wwwphpbb}.
A Bugzilla
\cite{wwwbugzilla}
is provided for 
bug tracking and feature requests.
The front end of \GPWR, i.e., the browser-based user interface, as well 
as the back-end routines interacting with the cluster software, 
are a PHP-based custom product developed by the \GP\ team.

\GPWR\ calculations submitted via the web interface are added to a run queue. 
If enough cores are available for computation, the 
calculation in the queue starts immediately. 
Otherwise, the queue manager schedules runs in such a way that 
each \WR\ user gets a fair share of resources, and available 
hardware is used at maximum capacity. 
Currently, all \GP\ v54 runs are assigned 12 cores
and up to 16 GB of memory per run, resulting in typical run 
times for a \GP\ calculation ranging from a few 
minutes (if only 2D cosmic-ray propagation is involved) to several hours
(if high resolution 3D cosmic ray/diffuse emission calculations are 
performed). 
The legacy \GP\ (v50) does not support multithreading,
and each calculation with this version of \GP\ is given 1 processor core. 
Our policy for allocation of computing resources may change in the future 
in order to better serve users interested in computationally intensive or 
batch runs.
Users are encouraged to provide feedback on their experience with \GPWR\ 
to the development team.

\section{Using the \GPWR\ service}

The interface of \WR\ consists of a narrow sidebar menu on the left and a wider main panel on the right of the screen (see Figure~\ref{webrun}). The links in the sidebar menu open in the main panel the interface sections `WebRun Help', `Configure \& Submit', `Monitor Queue', `Download Results' and `Exchange Runs', and the workflow of running and working with a calculation follows the sequence in which these sections are ordered. 

When users first log on and visit the \WR\ page, they see the `Configure \& Submit' section, in which a default GALPROP model is loaded. The default model does not correspond to any published results, but serves as a base with most parameters set to reasonable values. More models reproducing published results are provided as examples accessible via the drop-down menu at the top of the main panel.
To assist the user, in the `Configure \& Submit' section, \GP\ parameters are arranged into groups (the groups are navigable via scrolling or a shortcut bar at the top of the main panel), and some parameters are hidden to simplify navigation. The sidebar menu features options `First-time User Mode' and `Advanced User Mode', with more parameters visible in the latter. In both modes, some parameters are hidden or shown depending on values of other parameters. For example, when the user changes the \GP\ parameter `gamma\_rays' from its default value of `0' to `1', additional parameters in the `Gamma-Ray Emission' group appear, such as `skymap\_format', which was irrelevant for `gamma\_rays' equal to 0. To see all parameters, the user may choose the `Advanced User Mode'  and check the box `Show Inactive Parameters' in the sidebar menu. Some parameters, such as names of gas and dust maps, debugging-related and obsolete parameters, are fixed and cannot be edited in \WR, but can be perused by checking the box `Show fixed parameters'. At the bottom of the main panel, the button `Submit' saves the user-specified model parameters and sends them to a validation routine, which helps prevent misconfigured runs by providing detailed warning messages when parameters are out of range or have conflicting values. Upon a successful validation (or if the user chooses to ignore the warnings), the user will be prompted for a confirmation and given the option to download the formed GALDEF file (i.e., the \GP\ configuration file with specified parameter values). Detailed explanation of the meaning of all parameters is provided in the web form and in the GALPROP Explanatory Supplement \cite{wwwgalpropmanual}.

\begin{figure*}
   \centering
   \includegraphics[width=2\columnwidth]{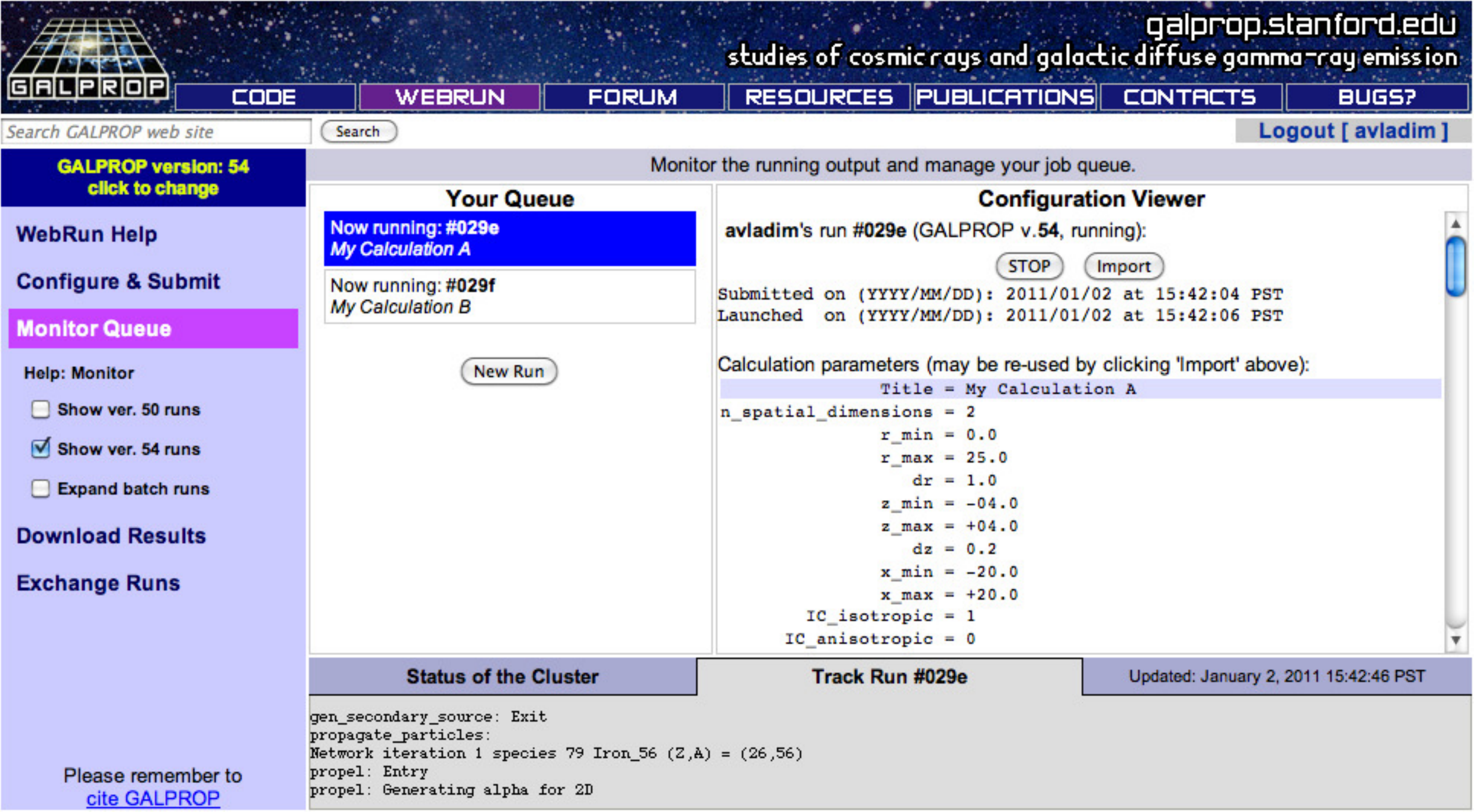}
   \caption{\GP\ calculations in \WR\ are executed on a dedicated computing cluster at Stanford University. A job queue with multiple simultaneous runs is supported. Each calculation is parallelized across several processor cores, and real time running output can be viewed via the \WR\ page.}   
   \label{monitor}
\end{figure*}

When the user confirms the run, it will be submitted to a run queue on the \GP\ computing cluster, and the user will be taken to the \WR\ section `Monitor Queue' (see Figure~\ref{monitor}). The main panel in this section contains: a list of queued and running calculations on the left; a configuration viewer panel on the right, which displays the configuration file and control buttons (`Stop' and `Import') for the run selected from the list; and a real-time monitor at the bottom, which displays the chosen calculation's running output or the status of the cluster. Parameter `Title', displayed along with the calculation number in the list, is very helpful in organizing and identifying one's runs. While the user may close the browser and/or log out from the \GP\ web site, it will not stop queued or running calculations. The runs will be finished, packaged and moved to the list in the `Download Results' section regardless of the user's online status.

\begin{figure*}
   \centering
   \includegraphics[width=2\columnwidth]{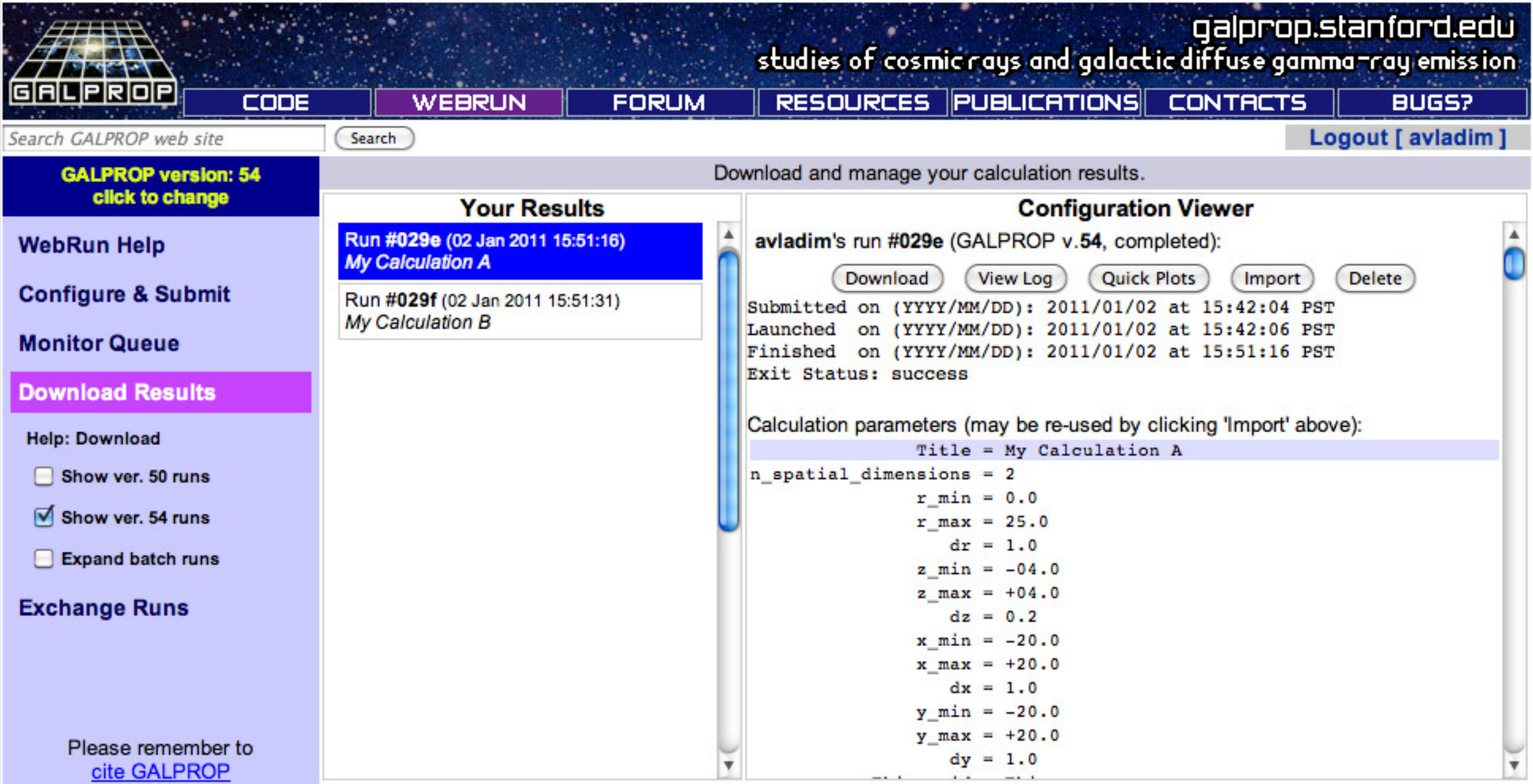}
   \caption{\GPWR\ users have access to and control over archives with results of all calculations they have performed. Each archive can be shared by publishing its URL. Visualization of results is possible within the \WR\ window via the `Quick Plots' function.}   
   \label{download}
\end{figure*}

\begin{figure*}[t]
   \centering
   \includegraphics[width=2\columnwidth]{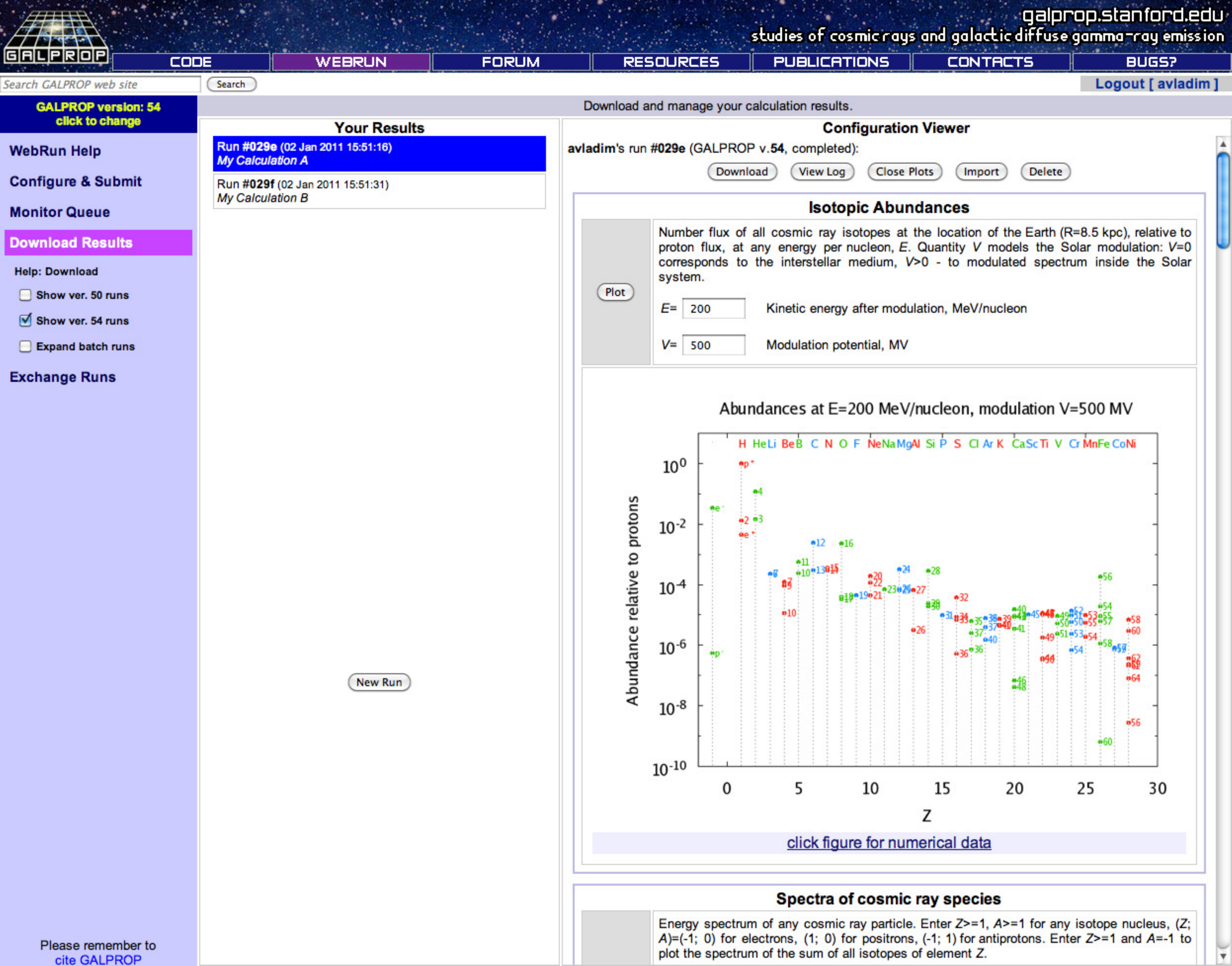}
   \caption{Quick Plots is an online plotting tool in \GPWR, capable of parsing the FITS files produced by GALPROP and creating raw text data tables and plots using built-in templates. Users are encouraged to develop their own plotting routines for publication purposes using the data produced by Quick Plots as a reference.}   
   \label{quickplots}
\end{figure*}

Completed runs may be viewed in section `Download Results', interface layout of which is similar to that of the `Monitor Queue' section, but there is no monitor panel (see Figure~\ref{download}). Clicking a run in the list displays its configuration and controls in the viewer panel: `Download', `View Log', `Quick Plots', `Import' and `Delete'. The `Download' button displays the link to the archive file with FITS files produced by \GP, and this link can be shared with the public, yet it is protected from unauthorized access by a random 16-character suffix. The runtime log produced by the code, included in the archive, can also be viewed online using the `View Log' button. `Quick Plots', a convenient tool illustrated in Figure~\ref{quickplots}, allows one to take a quick look at the results of the run. The `Import' button is useful for making a number of similar calculations. Clicking the button takes the user to the `Configure \& Submit' section, where the values of all parameters are identical to those in the imported run. Deleting the run using the `Delete' button irreversibly removes it from the run list and also erases it from the hard drive on the server, but the record of that run (user id, run id, timing information and exit status) is retained for usage statistics collection.

There are two more functions available in \WR : batch runs and run exchange. The link to batch runs can be found in the sidebar menu in the `Configure \& Submit' submenu. To execute a batch of runs, the user can submit an archive in {\it ZIP} or {\it tar} format with GALDEF files using this function. Run exchange is instrumental for collaborations, allowing any registered user to gain access to the log, Quick~Plots and Import capability of another user's run, as long as the link to the archive has been shared. Both of these functions are documented in the help pages and within the interface.

\section{Getting help and providing feedback}

The \GPWR\ service includes help pages and a FAQ 
that explains the features of the 
web interface to the \GP\ code. 
Information about the \GP\ code is available from the \GP\ 
web site
\cite{wwwgalpropmanual}.
A forum on the \GP\ web site
\cite{wwwgalpropforum}
allows registered \GP\ users to discuss the code, ask the developers 
additional questions, and provide feedback.

\section{How to acknowledge the use of GALPROP and WebRun}

If \GP\ is used to obtain results for your publication, please 
cite all \GP\ papers relevant to your results and the 
address of the \GP\ web site
\cite{wwwgalprop}.
In addition, if the \WR\ service is used, also cite the present 
paper, Vladimirov et al., 2010 in Computer Physics Communication. 
These terms may be updated in the future, with up-to-date instructions
in the `Terms of Use' section of
the \GP\ web site
\cite{wwwgalpropterms}.

\section{Acknowledgements}

The \GP\ code and \GPWR\ service are supported by NASA grant NNX09AC15G. 
The \GP\ project is also supported by the
Max-Planck Institut f\"{u}r extraterrestrische Physik and the Rechenzentrum
Garching (RZG) of the Max Planck Society and the IPP.
We also thank Advanced Micro Devices (AMD)
\cite{wwwamd}
and 
Colfax International
\cite{wwwcolfax}
for hardware and assistance with the computing cluster serving \WR.


\bibliographystyle{model1-num-names}
\bibliography{ms}







\end{document}